\documentclass[twocolumn,showpacs,amsmath,amssymb,prx,footinbib,floatfix,superscriptaddress]{revtex4-1}
\usepackage{graphicx}
\usepackage{savesym}
\usepackage{amsfonts}
\usepackage{bm}
\usepackage{color}
\usepackage{hyperref}
\usepackage{subfigure}
\usepackage{wasysym}
\usepackage{upgreek}
\usepackage{float}
\usepackage{dsfont}
\usepackage{mathtools}% Loads amsmath

\newcommand{\dagga}{{\phantom{\dagger}}}

\begin{document}

\title{Neural Gutzwiller-projected variational wave functions}

\author{Francesco Ferrari}
\email{frferra@sissa.it}
\affiliation{SISSA-International School for Advanced Studies, Via Bonomea 265, I-34136 Trieste, Italy}

\author{Federico Becca}
\affiliation{Dipartimento di Fisica, Universit\`a di Trieste, Strada Costiera 11, I-34151 Trieste, Italy}

\author{Juan Carrasquilla}
\affiliation{Vector Institute, MaRS Centre, Toronto, Ontario, M5G 1M1, Canada}
\affiliation{Department of Physics and Astronomy, University of Waterloo, Ontario, N2L 3G1, Canada}

\date{\today}

\begin{abstract}

Variational wave functions have enabled exceptional scientific breakthroughs related to the understanding of novel phases of matter. Examples include the 
Bardeen-Cooper-Schrieffer theory of superconductivity, the description of the fractional quantum Hall effect through the Laughlin state, and Feynman's 
variational understanding of large-scale quantum effects in liquid Helium. More recently, Gutzwiller-projected wave functions, typically constructed from 
fermionic degrees of freedom, have been employed to examine quantum spin models in the presence of competing interactions, where exotic phases with no 
spontaneous symmetry breaking and fractional excitations may exist. In this work, we investigate the aforementioned fermionic wave functions supplemented 
with neural networks, specifically with the so-called restricted Boltzmann machine (RBM), to boost their accuracy and obtain reliable approximations to 
the ground state of generic spin models. In particular, we apply our neural augmented fermionic construction to the description of both magnetically
ordered and disordered phases of increasing complexity, including cases where the ground state displays a non-trivial sign structure. Even though the RBM
state is by far more effective for N\'eel states endowed with a particularly simple sign structure, it provides a significant improvement over the original
fermionic state in highly frustrated regimes where a complex sign structure is anticipated, thus marking the path to an understanding of strongly-correlated
spin models on the lattice via  neural  Gutzwiller-projected variational wave functions.
\end{abstract}

\maketitle

\section{Introduction}

Fractionalization refers to a set of phenomena in quantum many-body physics where a collection of strongly interacting microscopic degrees of freedom,
such as electrons or spins on a lattice, break down into multiple quasiparticles. One of the most prominent examples is the fractional quantum Hall
effect~\cite{tsui1982}, where the constituent electrons decay into quasiparticles carrying fractions of the electron charge. Another prime example of
fractionalization arises in the study of spin liquids~\cite{balents2010,savary2016}, which are highly-entangled states of matter that host a wide array
of exotic phenomena such as emergent gauge structures, non-local excitations, and fractional quasiparticles. These quasiparticles, often termed ``spinons'',
behave as fractions of ordinary magnons and spin waves.

Traditionally, gauge theories have played a central role in the study of quantum spin liquids since their particular structure can capture the highly
peculiar non-local excitations and entanglement properties present in these exotic phases~\cite{balents2010,savary2016,wen2004}. To construct such gauge
theories, physicists rely on representations of the original degrees of freedom in terms of pairs of bosons or fermions, whose local Hilbert space is
constrained to a subspace corresponding to the original Hilbert space of the spin system. This reformulation results in a rich theory of quasiparticles
coupled to gauge fields with the potential to shed light onto the properties of the original system.

Since a gauge theoretical reformulation of a quantum spin system is formally exact, an explanation of all physically conceivable phases in the original
models is always possible. Whereas conventional phases of matter, such as N\'eel antiferromagnets and valence bond solids, are associated with the
confining phases of the gauge theory, fractionalization is associated with the deconfined phases where the spinons materialize as legitimate excitations
of the system. Thus, the hope is that this reformulation may help us understand exotic phases which naturally realize the quasiparticles and the emerging
gauge fields of the theory. Originally suggested by Baskaran and Anderson in the context of high temperature superconductivity~\cite{baskaran1988},
fractionalization and emergent gauge structures have been demonstrated in several exactly solvable systems such as Kitaev's toric code~\cite{kitaev2003} and 
honeycomb model~\cite{hermanns2018}, string-net condensates~\cite{levin2005}, as well confirmed in microscopic models~\cite{cincio2013,isakov2011,hickey2019},
variational wave functions~\cite{zhang2012}, and also experiments~\cite{tsui1982}.

In practice, however, gauge theories of quantum spin systems are typically intractable. In such cases, the generic procedure, also known as the ``parton''
construction, is to rewrite the original model in terms of canonical bosons or fermions, perform a mean-field decoupling of the gauge theory, and
then study the implications of the mean-field theory and its stability against fluctuations~\cite{wen2004}. Alternatively, a numerical approach
can be introduced, where wave functions from a mean-field treatment of the gauge theory are numerically projected back to the original Hilbert space
through the variational Monte Carlo method (VMC)~\cite{becca2017}.

The accuracy and degree of applicability of these states, here referred to as Gutzwiller projected wave functions, have been demonstrated in several
prototypical models of frustrated magnetism~\cite{sheng2009,iqbal2011,iqbal2013,gong2013,hu2015,iqbal2016}. The main advantage of this strategy is
that the states it produces can lead to very accurate estimates of the ground-state energy but their most important characteristic is the interpretability,
as these states often provide us with a physically transparent description of the phases of the model under consideration. On the other hand, the main
disadvantage of this approach is the lack of systematic ways of improving the quality of the approximations the construction introduces. While the parton
construction remains a powerful approach to quantum spin liquids, it is clear that the Gutzwiller projector only partially reintroduces the missing gauge fluctuations of the mean-field treatment,
which may not be sufficient to capture important long wavelength properties of the system. This effect is particularly important in special cases, e.g., in
$U(1)$ spin liquids, since Gutzwiller projected states do not account for crucial spatial gauge fluctuations~\cite{tay2011,liu2014}.

Here, we explore the possibility of improving the accuracy of the parton construction through neural networks. In particular, because of their extensive
use in condensed matter physics and quantum information~\cite{carleo2017,chen2018,torlai2018,torlai2018b,deng2017,gao2017,deng2017b,nomura2017,clark2018},
we consider restricted Boltzmann machines (RBM), which, due to the nature of their non-local structure, have been shown to represent some highly-entangled
many-body states using a relatively small number of parameters~\cite{deng2017,gao2017,chen2018,glasser2018}. The family of states we consider can be thought
of as quantum mechanical version of the so-called product of experts idea used in probabilistic models in machine learning~\cite{hinton2002}, where the core
strategy is to combine several simpler wave functions (the ``experts'') by multiplying their amplitudes in a certain computational
basis~\cite{becca2017,toulouse2008,carrasquilla2013,nomura2017,glasser2018}. Concretely, the family of states we consider is written as a product of a
Gutzwiller-projected fermionic state and a complex-valued RBM. By virtue of the high representation power of the RBM~\cite{leroux2008,gao2017}, the expectation
is that the RBM may help us improve the accuracy of the Gutzwiller projected states and in part alleviate the problem of the missing gauge fluctuations in these
family of states.

The paper is organized as follow: In Sec.~\ref{sec2}, we introduce the spin models that are considered in this study; in Sec.~\ref{sec3}, we define the 
variational wave functions; in Sec.~\ref{sec4}, we briefly describe the numerical optimization of the wave functions; in Sec.~\ref{sec5}, we discuss the results; finally, in 
Sec.~\ref{sec6}, we draw our conclusions.

\section{Spin models}\label{sec2}

We consider two-dimensional (2D) quantum lattice models that are traditionally studied in condensed matter physics. These models host a wide array of
effects that are relevant to the understanding of physical phenomena such as low-temperature experiments with superfluid helium~\cite{daunt1954},
Fe-based superconducting materials~\cite{kamihara2008}, triangular lattice compounds~\cite{paddison2017}, fractional excitations in the quantum
antiferromagnets~\cite{dallapiazza2015}, among many others. We begin by introducing the $J_1-J_2$ Heisenberg model, an archetypical model of frustration
in quantum antiferromagnets. Its Hamiltonian is given by
\begin{equation}\label{eq:j1j2model}
 \mathcal{H}=J_{1} \sum_{\langle i,j \rangle}  \mathbf{S}_i \cdot \mathbf{S}_j +
 J_{2} \sum_{\langle\langle i,j \rangle\rangle}  \mathbf{S}_i \cdot \mathbf{S}_j,
\end{equation}
where $\mathbf{S}_i=(S^x_i,S^y_i,S^z_i)$ are spin-1/2 degrees of freedom sitting on the sites of a square lattice and $J_1,J_2\geq 0$ characterize the
antiferromagnetic couplings between the magnetic degrees of freedom. Here, $\langle \cdots \rangle$ and $\langle \langle \cdots \rangle \rangle$ restrict
the summations over first- and second-nearest neighboring sites, respectively. Throughout this paper, we focus on two different values of the
{\it frustrating ratio}, namely $J_2/J_1=0$ and $J_2/J_1=0.5$, to assess the accuracy of the aforementioned variational wave functions in two different
regimes. For $J_2=0$ the model reduces to the (unfrustrated) Heisenberg model, whose ground state displays long-range N\'eel magnetic
order~\cite{sandvik1997,calandra1998}, At finite $J_2$, the energetic competition between the antiferromagnetic couplings at first- and second-nearest
distances introduces frustration, resulting in a reduction of the magnetic order. In particular, the system exhibits a high degree of frustration around
$J_2/J_1\approx 0.5$, where the nature of the ground state of the model is a long standing problem~\cite{read1989,schulz1996}. Different scenarios have
been suggested, such as the onset of a valence-bond solid phase, with columnar~\cite{poilblanc2017,haghshenas2018} or plaquette~\cite{mambrini2006,gong2014}
order, and/or the existence of a spin-liquid phase~\cite{hu2013,morita2015,sandvik2018}.

Due to its relevance to physical situations ranging from low-temperature experiments with superfluid helium~\cite{daunt1954}, magnetic
insulators~\cite{giamarchi2008}, and ultracold gases in optical lattices~\cite{bloch2008,carrasquilla2012}, we also consider the $XY$ model on the square
lattice, whose Hamiltonian is given by
\begin{equation}\label{eq:xymodel}
 \mathcal{H}=J \sum_{\langle i,j \rangle}  (S^x_i S^x_j + S^y_i S^y_j).
\end{equation}
In this case, the ground state exhibits N\'eel order with the spins aligned in the $xy$ plane. A fundamental difference between the XY and the unfrustrated
Heinseberg models comes from their symmetries: while the Heisenberg exchange interaction is invariant under global spin rotations around any axis (i.e.,
it exhibits $SU(2)$ symmetry), the $XY$ coupling is invariant only under global rotations around $S_z$ (leading to a $U(1)$ symmetry).

Finally, we consider the Heisenberg model on the triangular lattice, a model of geometric frustration relevant to the understanding of organic Mott
insulators displaying spin liquid behavior~\cite{shimizu2003}. Its Hamiltonian is given by Eq.~(\ref{eq:j1j2model}) with $J_2=0$. Due to frustration,
this system exhibits strong quantum fluctuations. Even though this model was initially proposed as a candidate host for the resonating valence-bond spin
liquid~\cite{anderson1973}, which motivated the entire field of spin liquids, numerical evidence~\cite{capriotti1999,white2007} suggested that the ground
state of the system is magnetically ordered with a $120^\circ$ pattern, which is the accepted consensus.

\section{Construction of the wave functions}\label{sec3}

In general, the variational {\it Ans\"atze} employed in this work can be written as a product of a Gutzwiller-projected fermionic wave function,
$|\Psi_{f}\rangle$, and a many-body correlator,
$\hat{\mathcal{C}}$:
\begin{equation}\label{eq:fullwf}
 |\Psi_{\mathcal{C}}\rangle=\hat{\mathcal{C}} |\Psi_{f}\rangle
 =\sum_{\sigma} \mathcal{C}(\sigma) \langle \sigma|\Psi_{f}\rangle |\sigma\rangle.
\end{equation}
In the above formula we have inserted a resolution of the identity ($\hat{\mathds{1}}=\sum_{\sigma}|\sigma\rangle\langle\sigma|$) and exploited the fact
that  $\hat{\mathcal{C}}$ is diagonal in the many-body computational basis $\{|\sigma\rangle\}$.

In order to construct the fermionic part of the {\it Ansatz} we turn to a parton construction, representing the spin degrees of freedom in terms of
Abrikosov fermions~\cite{wen2004}:
\begin{equation}\label{eq:abrikosov}
\mathbf{S}_i=\frac{1}{2}\sum_{\alpha,\beta} f^\dagger_{i,\alpha} \boldsymbol{\sigma}_{\alpha,\beta} f_{i,\beta}.
\end{equation}
Here $\boldsymbol{\sigma}=(\sigma^x,\sigma^y,\sigma^z)$ is the vector of Pauli matrices and the operator $f^\dagger_{i,\alpha}$ ($f^\dagga_{i,\alpha}$)
creates (annihilates) a fermion with spin $\alpha$ at site $i$. Within the Abrikosov picture, the commutation relations among spins are a consequence of
the fermionic anticommutation relations. However, the fermionic representation of spins enlarges the Hilbert space of the system, allowing each lattice
site to be in four different states, namely $|0\rangle$, $|\uparrow\rangle$, $|\downarrow\rangle$, and $|\uparrow\downarrow\rangle$. Therefore, the
mapping of Eq.~(\ref{eq:abrikosov}) is a faithful representation of spin degrees of freedom only if the fermions are restricted to the subspace of
configurations with one fermion per site. This constraint can be enforced by the Gutzwiller projection operator
$\mathcal{P}_G=\prod_i (n_{i,\uparrow}-n_{i,\downarrow})^2$ (where $n_{i,\sigma}=f^\dagger_{i,\sigma}f^\dagga_{i,\sigma}$). The wave function
$|\Psi_{f}\rangle$ is obtained by Gutzwiller-projecting a certain fermionic state $|\Phi_0\rangle$ (see below):
\begin{equation}
 |\Psi_{f}\rangle= \mathcal{P}_{S^z_{tot}=0}\mathcal{P}_G |\Phi_0\rangle.
\end{equation}
In addition to the Gutzwiller projector, here we consider a second projector, $\mathcal{P}_{S^z_{tot}=0}$, which restricts the wave function to the
subspace of configurations with ${S^z_{tot}=\sum_i S^z_i=0}$. We emphasize the fact that both projections are treated exactly by sampling only the
fermionic configurations which satisfy the desired constraints. Finally, $|\Phi_0\rangle$ is obtained by computing the ground state of a quadratic
(i.e., a mean-field) Hamiltonian ($\mathcal{H}_0$) of Abrikosov fermions,
\begin{equation}\label{eq:H0}
 {\cal H}_{0} = {\cal H}_{\rm BCS} + {\cal H}_{\rm AF},
\end{equation}
which we have split into two parts for clarity. The first part is a Bardeen-Cooper-Schrieffer (BCS) Hamiltonian containing a (complex) hopping term
($t_{i,j}$) and a singlet pairing term ($\Delta_{i,j}=\Delta_{j,i}$):
\begin{equation}\label{eqn:BCS}
   {\cal H}_{\rm BCS} = \sum_{i,j,\sigma} t_{i,j} f_{i,\sigma}^\dagger f_{j,\sigma}^\dagga
              + \sum_{i,j} \Delta_{i,j} f_{i,\uparrow}^\dagger f_{j,\downarrow}^\dagger + h.c..
\end{equation}
The above Hamiltonian is invariant under global spin rotations and, therefore, the spin wave functions obtained by the Gutzwiller projection of its
ground states are $SU(2)$ symmetric. {\it Ans\"atze} of these form are particularly suited to describe magnetically disordered phases of matter, such as
quantum spin liquids and valence-bond solids~\cite{savary2016}. The Gutzwiller-projected BCS wave functions fall back into the class of resonating
valence-bond states, first introduced by Anderson in the context of high-Tc superconductivity~\cite{anderson1987}. The second term of Eq.~(\ref{eq:H0}),
${\cal H}_{\rm AF}$, contains a magnetic field in the $xy$-plane which induces magnetic order in the variational {\it Ansatz}:
\begin{equation}\label{eqn:AF}
    {\cal H}_{\rm AF} = \Delta_{\rm AF} \sum_{i}
             \left ( e^{i Q \cdot R_i} f_{i,\uparrow}^\dagger f_{i,\downarrow}^\dagga
            + e^{-i Q \cdot R_i} f_{i,\downarrow}^\dagger f_{i,\uparrow}^\dagga \right ).
\end{equation}
The vector $Q$ determines the periodicity of the magnetic order: the N\'eel phase on the square lattice is obtained by setting $Q=(\pi,\pi)$, while the
$120^\circ$ phase on the triangular lattice corresponds to $Q=(\frac{4\pi}{3},0)$ [or, equivalently, $Q=(\frac{2\pi}{3},\frac{2\pi}{\sqrt{3}})$].
Typically, the wave function obtained by Gutzwiller-projecting the ground state of ${\cal H}_{\rm AF}$ overestimates the magnetic order
parameter~\cite{becca2011}. To improve this construction, further quantum fluctuations can be added by including hopping terms in the BCS Hamiltonian and
by applying a two-body Jastrow factor correlator to the fermionic wave function [as in Eq.~(\ref{eq:fullwf})]:
\begin{equation}
 \mathcal{C}_{\mathrm{Jastrow}}(\sigma)=\exp\left(\frac{1}{2}\sum_{i,j} v_{i,j} \sigma^z_i \sigma^z_j\right).
\end{equation}
The Jastrow factor has the effect of adding correlations which are perpendicular to the in-plane magnetic field $\Delta_{\rm {AF}}$. In this work, we
consider Jastrow factors with translationally invariant long-range {\it pseudopotentials} ${v_{i,j}=v(|R_i-R_j|)\in \mathbb{R}}$. We emphasize the fact
that all the coupling constants defining ${\cal H}_{0}$ (i.e. $t_{i,j},\Delta_{i,j},\Delta_{\rm {AF}}$) and the Jastrow pseudopotentials play the role
of variational parameters, which are optimized in order to find the best variational energy. The optimal parametrization of the auxiliary Hamiltonian
${\cal H}_{0}$ depends on the model under investigation.

The central aim of this work is to explore the possibility of improving Gutzwiller-projected fermionic wave functions by applying a stronger many-body
correlator than the two-body Jastrow factor. For this purpose, a {\it neural network} is employed in the form of a restricted Boltzmann machine. This
network is defined by introducing a set of auxiliary Ising variables, $\{h^\alpha\}_{\alpha=1,...,N_\alpha}$, which form the so-called {\it hidden} layer.
These variables are coupled to the z-components of the spins of the lattice (dubbed as {\it visible} layer, $\{\sigma_i^z\}_{i=1,...,N}$) through a
classical energy functional of the form:
\begin{equation}\label{eq:HRBM}
 E_{\mathrm{RBM}}= \sum_{i=1}^N \sum_{\alpha=1}^{N_\alpha} h^\alpha W^\alpha_i \sigma^z_i +
 \sum_{\alpha=1}^{N_\alpha} b^\alpha h^\alpha + \sum_{i=1}^N a_i \sigma^z_i.
\end{equation}
The RBM correlator is then obtained by computing the Boltzmann factor $e^{E_{\mathrm{RBM}}}$ and taking its trace over the hidden variables degrees of
freedom. This operation can be performed exactly due to the particular form of the classical energy functional of Eq.~(\ref{eq:HRBM}), which only contains
interactions between variables belonging to the two different layers (i.e., no intralayer couplings are allowed). Since we consider translationally
invariant states with conserved magnetization $S^{z}_{tot}=0$, we can set $a_i=0$ and obtain the final form of the RBM correlator~\cite{carleo2017}:
\begin{equation}\label{eq:rbm_nosymm}
 \mathcal{C}_{\mathrm{RBM}}(\sigma)=\exp \left[  \sum_\alpha \log\cosh\left(b^\alpha +
 \sum_i W^\alpha_i  \sigma^z_{i}   \right) \right].
\end{equation}
The parameters of $\hat{\mathcal{C}}_{\mathrm{RBM}}$ are called {\it biases} ($b^\alpha$) and {\it weights} ($W^\alpha_i$), and are (in general) assumed
to be complex numbers in this work. The complex parametrization of the RBM allows the correlator to change both the amplitudes and the phases of the
fermionic wave function to which it is applied. The expression of Eq.~(\ref{eq:rbm_nosymm}) can be regarded as a sort of many-body Jastrow factor, since
a series expansion of the $\log\cosh(\dots)$ function contains the n-body terms of the $\sigma^z$ variables. Unlike the Jastrow factor, the RBM correlator
not only breaks the $SU(2)$ symmetry of spin, but also the $Z_2$ symmetry $\sigma^z \mapsto -\sigma^z$. This happens if the biases $b^\alpha$ are nonzero,
since the aforementioned expansion can contain products of odd numbers of spins.

An important question to address is the implementation of lattice symmetries in the correlator $\hat{\mathcal{C}}$. As already pointed out, a symmetric
two-body Jastrow factor can be obtained by simply taking a symmetric pseudopotential, e.g., $v_{i,j}=v(|R_i-R_j|)$. This procedure, however, cannot be
applied to the RBM correlator, since its parameters depend upon the index $\alpha$ (labelling the hidden units), which does not have any physical meaning.
Therefore, the most straightforward way of including symmetries is to implement them {\it a posteriori}. Concretely, if we want to enforce translational
symmetry, we symmetrize the RBM correlator through a product over all possibile Bravais lattice translations $\{\mathcal{T}_R\}$:
\begin{align}\label{eq:rbm_translations}
 & \mathcal{C}_{\mathrm{tRBM}}(\sigma)=\prod_{R} C_{\mathrm{RBM}}[\mathcal{T}_R(\sigma)] \nonumber \\
 &=\exp \left[ \sum_{R} \sum_\alpha
 \log\cosh\left(b^\alpha +  \sum_i W^\alpha_i  \sigma^z_{i+_R}   \right) \right].
\end{align}
The above expression is translationally invariant with momentum $K=(0,0)$. In addition, the point group symmetries $\{\Sigma\}$ of the lattice can be
implemented on top of the translationally invariant correlator. The procedure is the same as the one employed in Eq.~(\ref{eq:rbm_translations}):
\begin{align}\label{eq:rbm_allsymm}
 & \mathcal{C}_{\mathrm{sRBM}}(\sigma)=
 \prod_{\Sigma}\prod_{R} C_{\mathrm{RBM}}[\Sigma\mathcal{T}_R(\sigma)] \nonumber \\
 &=\exp \left[ \sum_{\Sigma}\sum_{R} \sum_\alpha
 \log\cosh\left(b^\alpha +  \sum_i W^\alpha_i  \sigma^z_{\Sigma(i+_R)}   \right) \right],
\end{align}
where $\Sigma(j)$ indicates the position of the site obtained by applying the symmetry $\Sigma$ to the site $j$. The quantum numbers of the above
correlators, which are associated to the different point group symmetries, are all zero by construction. A more general strategy to implement symmetries
with the desired quantum numbers is outlined in Ref.~\onlinecite{choo2018}.

The main advantage of the RBM correlator with respect to the Jastrow factor comes from the fact that its accuracy can be, in principle, systematically
improved by increasing the number of hidden variables $N_\alpha$. This is due to the fact that a RBM is a universal function approximator~\cite{leroux2008},
which means that it can approximate any function with arbitrary accuracy if the number of variables in the hidden layer is allowed to grow arbitarily large.
In addition, the non-local structure of this neural network makes it capable of capturing highly-entangled phases of matter. On the other hand, the
disadvantage of employing the RBM correlator in relation to the simpler two-body Jastrow factor mainly resides in its higher computational cost. Crucial for
an efficient variational Monte Carlo, computing ratios of translationally invariant RBMs has a cost which scales linearly with the number of sites and the
number of hidden units ($O(N\times N_\alpha)$), while computing ratios of two-body Jastrow factors simply scales as $O(1)$~\cite{becca2017}. Finally, another
disadvantage of the RBM is the lack of straightforward physical interpretability of its variational parameters, which are associated to many-body spin-spin
correlations at all distances. Instead, the pseudopotential $v_{i,j}$ of the Jastrow factor clearly accounts for the two-body correlation of spins $i$ and
$j$, and typically shows a clear physical behavior~\cite{capello2005}, decaying with the distance $|R_i-R_j|$.

\section{Optimization of the variational wave function}\label{sec4}

In order to minimize the variational energy of the wave function~\eqref{eq:fullwf}, namely
\begin{equation}
 E_\mathcal{C}=\langle \mathcal{H} \rangle_\mathcal{C}=
 \frac{\langle \Psi_\mathcal{C}| \mathcal{H} |\Psi_\mathcal{C} \rangle}{\langle \Psi_\mathcal{C}|\Psi_\mathcal{C} \rangle},
\end{equation}
we employ the stochastic reconfiguration (SR) technique~\cite{sorella2005,becca2017}, which is briefly summarized below.

Let us denote by $\omega=\{\omega_k\}$ the set of all the variational parameters of the RBM-fermionic state, which is formed by the couplings included in 
the auxiliary Hamiltonian $\mathcal{H}_0$ and by the weights and biases of the RBM correlator. For each parameter $\omega_k$, we can define a corresponding 
operator $\hat{\mathcal{O}}_k$ that is diagonal in the basis of spin configurations, i.e., 
${\langle \sigma|\hat{\mathcal{O}}_k|\sigma'\rangle= \mathcal{O}_k(\sigma)\delta_{\sigma,\sigma'}}$, and yields the logarithmic derivative of the amplitudes 
of $|\Psi_\mathcal{C}\rangle$:
\begin{equation}
 \mathcal{O}_k(\sigma)=
 \frac{\partial \log\left[\langle \sigma | \Psi_\mathcal{C}\rangle \right]}{\partial \omega_k}=
 \frac{1}{\langle \sigma | \Psi_\mathcal{C}\rangle }
 \frac{\partial \langle \sigma | \Psi_\mathcal{C}\rangle }{\partial \omega_k}.
\end{equation}
Let us also introduce another diagonal operator, the so-called \textit{local energy} $\hat{E}_{\rm loc}$, whose matrix elements read
\begin{equation}
\langle \sigma | \hat{E}_{\rm loc} | \sigma' \rangle =E_{\rm loc}(\sigma)\delta_{\sigma,\sigma'}
=\frac{\langle \sigma | \mathcal{H} | \Psi_\mathcal{C}\rangle}{\langle \sigma | \Psi_\mathcal{C}\rangle}
\delta_{\sigma,\sigma'}.
\end{equation}

At each step of the SR algorithm, the parameters of the variational wave functions are updated according to
\begin{equation}\label{eq:update_par}
 \omega_k' = \omega_k + \eta \sum_{k'} \mathcal{S}^{-1}_{k,k'} f_{k'}.
\end{equation}
Here $\eta$ is an arbitrary hyperparameter dubbed as the \textit{learning rate}, $\mathcal{S}^{-1}$ is the inverse of the \textit{covariance matrix}
\begin{equation}\label{eq:Smatrix_def}
 \mathcal{S}_{k,k'}=\Re \left[
 \langle \hat{\mathcal{O}}_k^\dagger  \hat{\mathcal{O}}_{k'}^\dagga \rangle_\mathcal{C}
 - \langle \hat{\mathcal{O}}_k^\dagger \rangle_\mathcal{C} \langle \hat{\mathcal{O}}_{k'}^\dagga\rangle_\mathcal{C}
 \right],
\end{equation}
and $f_k$ are the \textit{forces}~\cite{becca2017}
\begin{equation}
f_k= -\frac{\partial E_\mathcal{C}}{\partial \omega_k} = -2 \Re \left[ 
\langle \mathcal{H} \hat{\mathcal{O}}_k \rangle_\mathcal{C} -
\langle \mathcal{H} \rangle_\mathcal{C} \langle \hat{\mathcal{O}}_k \rangle_\mathcal{C}
\right].
\end{equation}
Both the elements of the covariance matrix and the forces are computed stochastically by Monte Carlo sampling (through the Metropolis 
algorithm~\cite{metropolis1953}), and then employed to update the variational parameters as in Eq.~\eqref{eq:update_par}. We note that a redundant 
parametrization of the variational wave function causes the covariance matrix to be non-invertible. This situation is commonly encountered when neural 
network correlators with a large number of variational parameters are considered. To avoid numerical instabilities connected to the singularity of the 
covariance matrix, we apply an explicit regularization of the form $\mathcal{S}_{k,k'}\mapsto (1+\epsilon \delta_{k,k'}) \mathcal{S}_{k,k'}$, where 
$\epsilon$ is an arbitary small parameter~\cite{becca2017}. 

For most of the variational results presented in this work, we performed few distinct optimizations starting from different set of initial parameters, 
and we selected as optimal wave function the one which provided the lowest variational energy. In general, we used $\approx 2\times 10^4$ up to 
$\approx 2 \times 10^5$ Monte Carlo samples to evaluate the forces and the covariance matrix of the SR method. We typically chose a learning rate $\eta$ 
in the range $[0.02,0.05]$ and a regularization parameter $\epsilon= 10^{-3}$. For what concerns the latter, in the case of variational wave functions 
with a large number of variational parameters (e.g. $N_\alpha=8$ on a $10\times 10$ lattice), we observed that starting the optimization with a larger 
value of $\epsilon$, e.g. $10^{-2}$, helps reducing some numerical instabilities and yields a smoother decrease of the variational energy; then, after 
few hundreds SR steps, the value of $\epsilon$ can be safely changed to $10^{-3}$ for the rest of the optimization.

\section{Results}\label{sec5}

%%%%%%%%%%%%%%%%%%%%%%%%%%%%%%%%%%%%%%%%%%%%%%%%%%%%%%%%%%%%%%%%%%%%%
\begin{figure}
\includegraphics[width=\columnwidth]{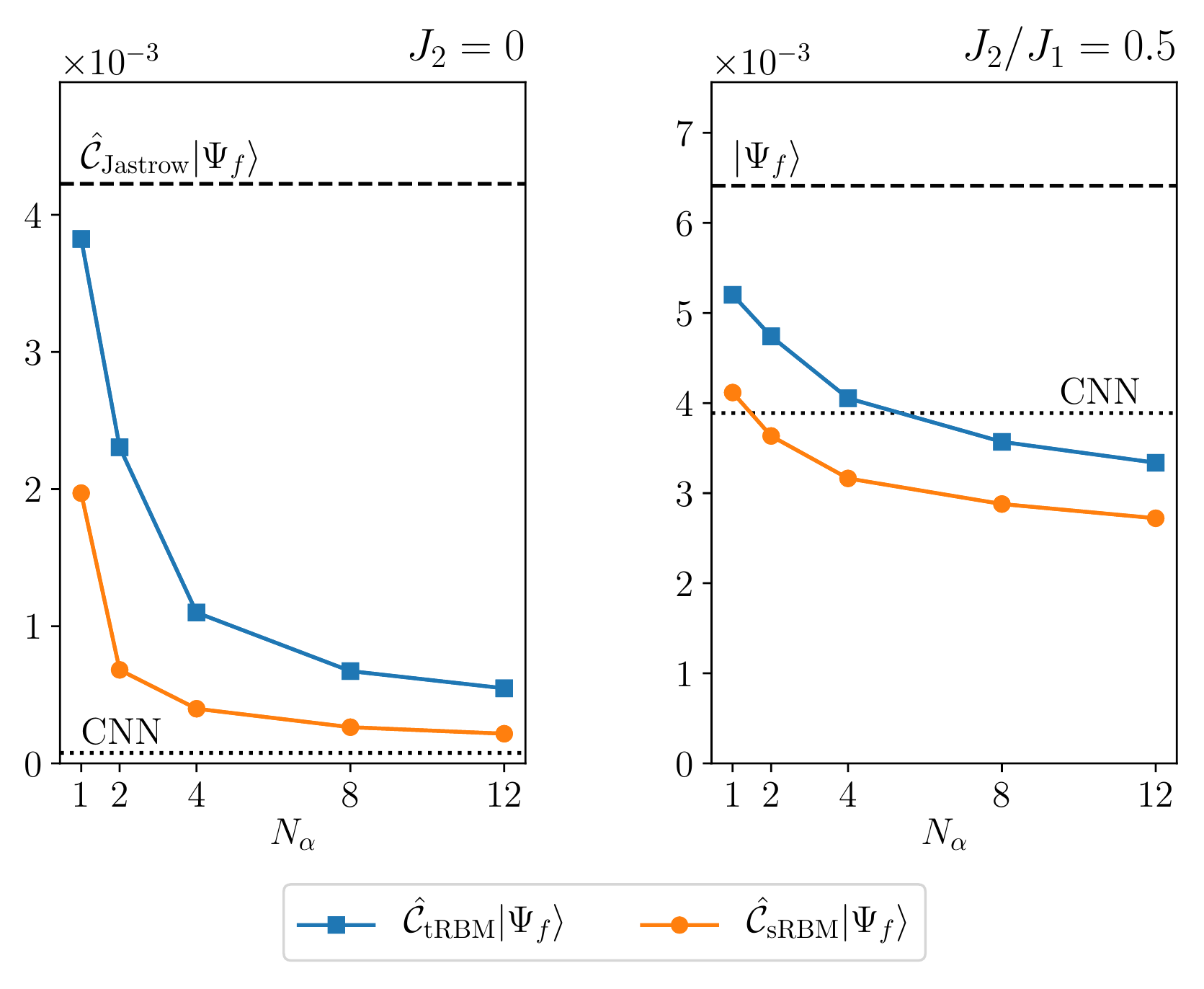}
\caption{\label{fig:energy_j1j2_6x6}
Relative error of the VMC energies $\Delta E$ [see Eq.~\eqref{eq:rel_err_energy}] with respect to the exact ones for the $J_1-J_2$ model on the $6\times 6$ 
square lattice. The results for the unfrustrated case ($J_2=0$) and the frustrated one ($J_2/J_1=0.5$) are shown on the left and on the right panel, 
respectively. The relative error of the RBM-fermionic wave function is plotted as a function of the number of hidden units: blue squares refer to the case 
of translationally invariant RBM correlator $\hat{\mathcal{C}}_{\mathrm{tRBM}}$, while orange circles correspond to the fully symmetric RBM correlator
$\hat{\mathcal{C}}_{\mathrm{sRBM}}$. The error-bars are smaller than the size of the dots. The dashed line represents the relative error of the
fermionic wave function of reference, which includes a Jastrow factor in unfrustrated case ($J_2=0$). The dotted line refers to the relative error
of the CNN quantum state of Ref.~\onlinecite{choo2019}.}
\end{figure}

\begin{figure}
\includegraphics[width=\columnwidth]{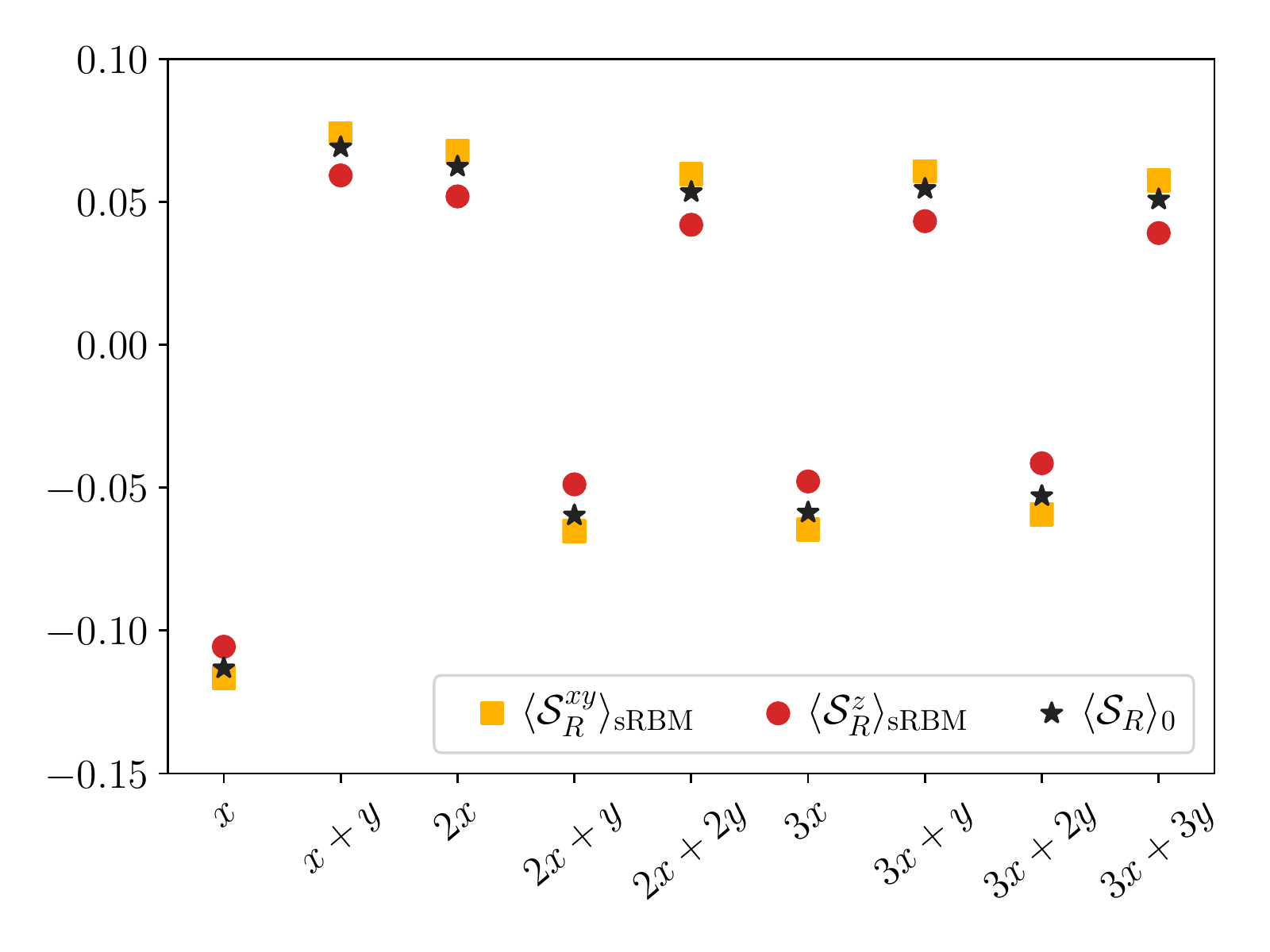}
\caption{\label{fig:corr_heis_6x6_r}
Spin-spin correlations for the Heisenberg model ($J_2=0$) on the $6\times 6$ square lattice, as a function of distance $R$. Here, $x=(1,0)$ and
$y=(0,1)$ are the lattice unit vectors. The set of data represented with yellow squares (red circles) corresponds to the expectation value of
$\mathcal{S}^{xy}_R$ ($\mathcal{S}^{z}_R$) on the RBM-fermionic wave function with the fully symmetric correlator $\hat{\mathcal{C}}_{\mathrm{sRBM}}$
($N_\alpha=1$). The error-bars are smaller than the size of the dots. The black stars represent the exact value of the spin-spin correlation, i.e.,
$\langle\mathcal{S}_R\rangle_0=1/3 \langle \mathbf{S}_0 \cdot \mathbf{S}_R \rangle_0$.}
\end{figure}

\begin{figure*}
\includegraphics[width=2\columnwidth]{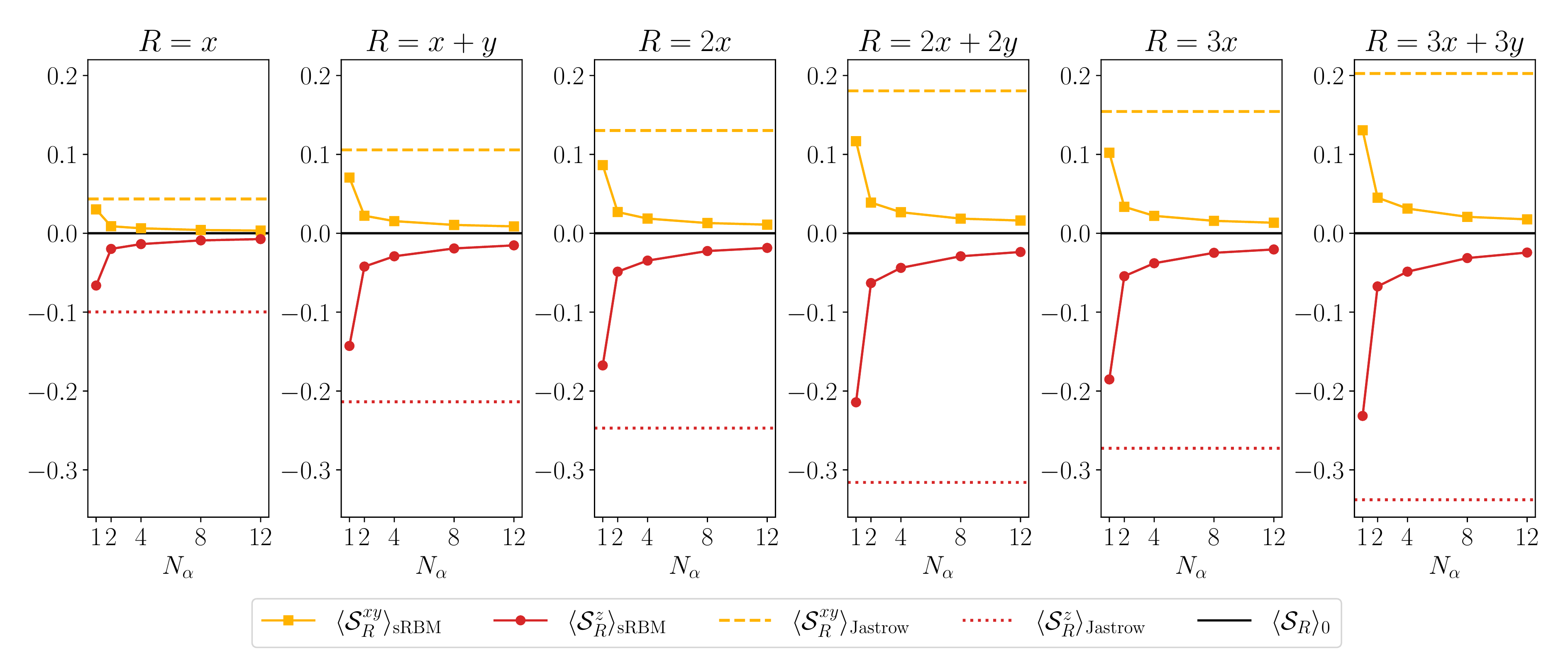}
\caption{\label{fig:corr_heis_6x6}
Relative error of the spin-spin correlations [see Eq.~\eqref{eq:rel_err_spinspin}] for the Heinseberg model ($J_2=0$) on the $6\times 6$ square lattice, 
as a function of the number of hidden units. The correlations are evaluated at different distances $R$, which are expressed in terms of the lattice unit 
vectors $x=(1,0)$ and $y=(0,1)$. The set of data represented with yellow squares (red circles) corresponds to the relative error of the expectation value 
of $\mathcal{S}^{xy}_R$ ($\mathcal{S}^{z}_R$), computed by employing the RBM-fermionic wave function with the fully symmetric correlator
$\hat{\mathcal{C}}_{\mathrm{sRBM}}$. The yellow dashed (red dotted) line, instead, refers to the relative error of the expectation value of
$\mathcal{S}^{xy}_R$ ($\mathcal{S}^{z}_R$) computed by employing the Jastrow correlator instead of the RBM. Finally, the black line indicates the zero
of the vertical axis, i.e.  the position of the exact value in the relative error scale.}
\end{figure*}

\begin{figure*}
\includegraphics[width=2\columnwidth]{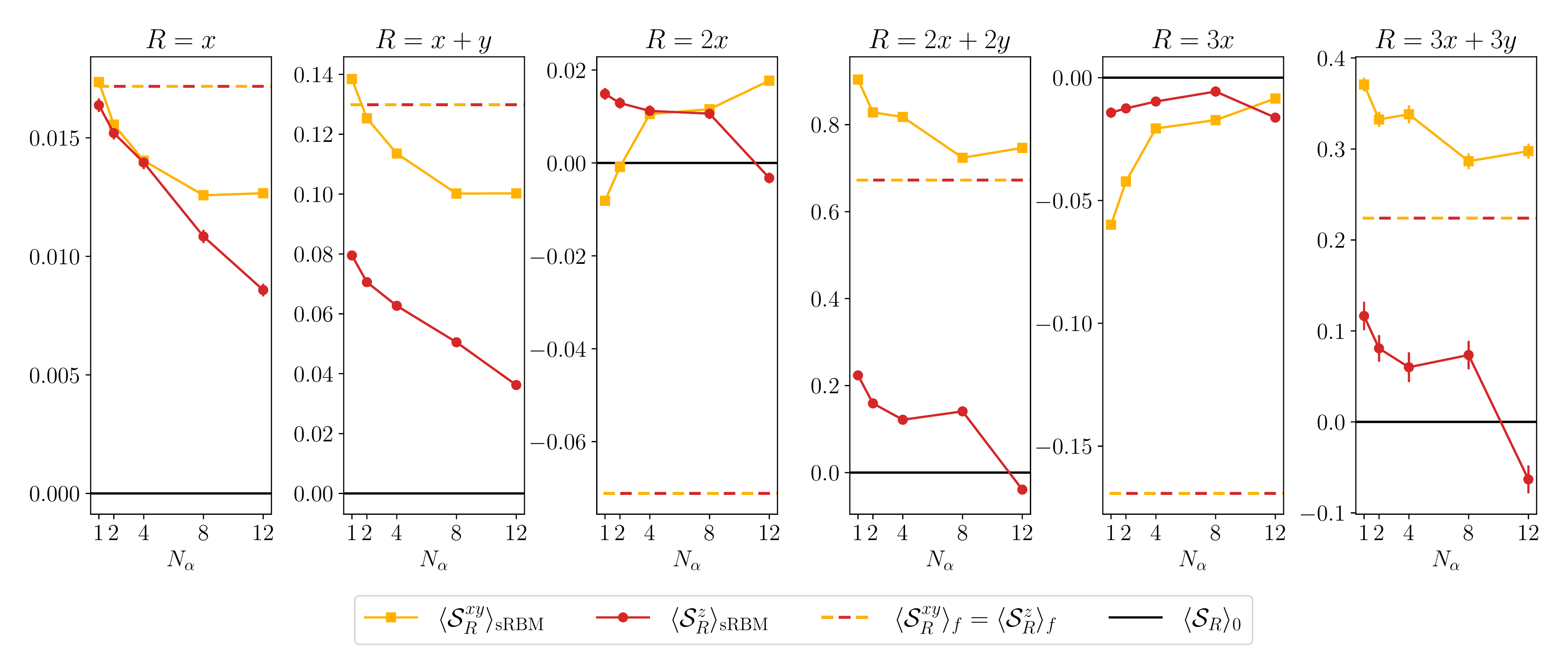}
\caption{\label{fig:corr_j2_6x6}
The same as in Fig.~\ref{fig:corr_heis_6x6} for $J_2/J_1=0.5$. The only difference is given by the fact that the fermionic state (without Jastrow factor)
is $SU(2)$ invariant and, therefore, in-plane and out-of-plane correlations are equal (and denoted by the bicolor dashed line).}
\end{figure*}

\begin{figure}
\includegraphics[width=\columnwidth]{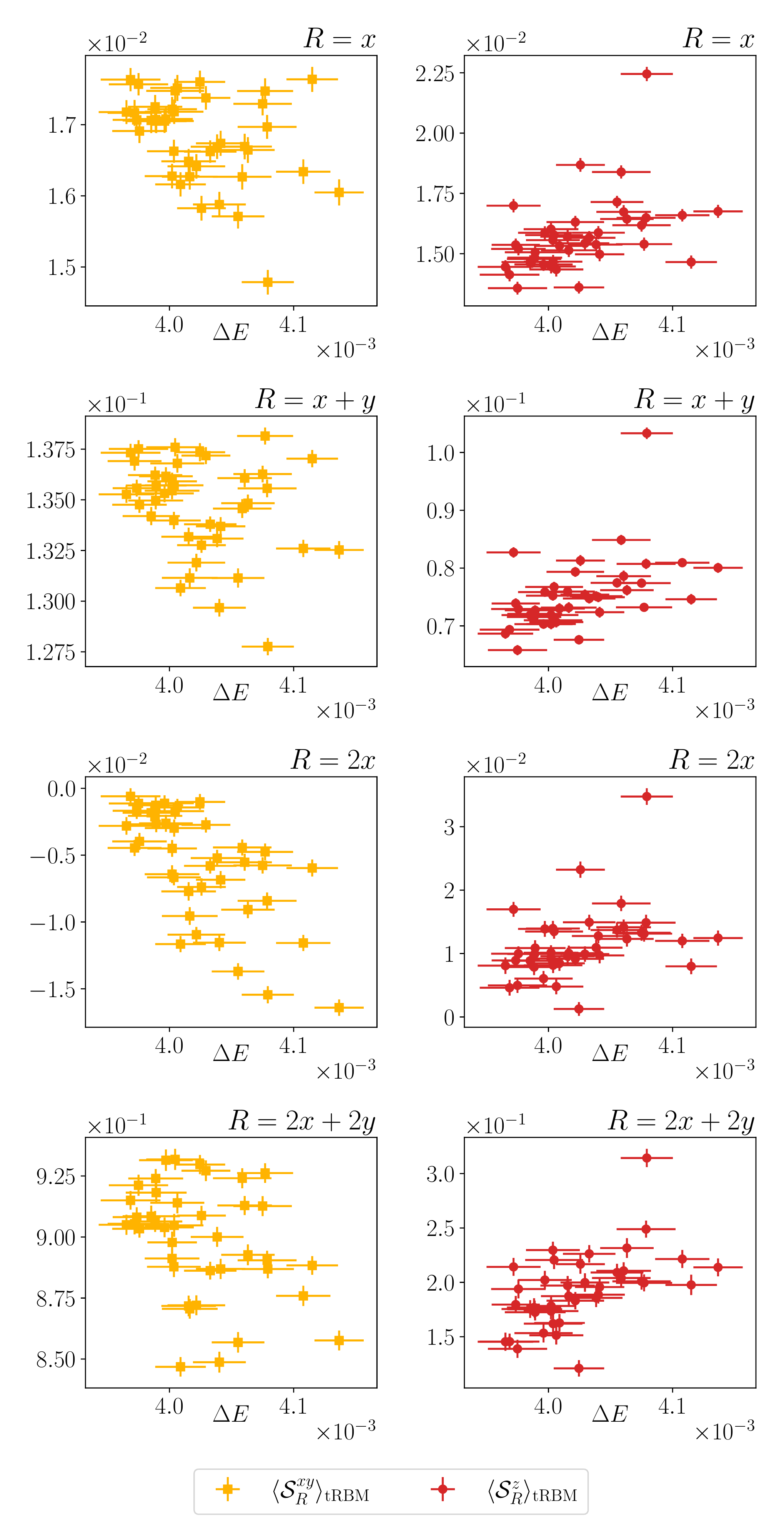}
\caption{\label{fig:scattered_corr}
Relative error of the spin-spin correlations [see Eq.~\eqref{eq:rel_err_spinspin}] as a function of the relative error of the variational energy for the 
$J_1-J_2$ model on the $6\times 6$ square lattice in the frustrated regime, $J_2/J_1=0.5$. The correlations are computed at different distances $R$, which 
are expressed in terms of the lattice unit vectors $x=(1,0)$ and $y=(0,1)$. The wave function employed in the calculations is a RBM-fermionic {\it Ansatz} 
with a translational invariant correlator ($N_\alpha=4$). Different points correspond to the results of different optimizations of the variational parameters. 
The set of data represented with yellow squares (red circles) corresponds to the relative error of the expectation value of the in-plane (out-of-plane)
correlations.}
\end{figure}

\begin{figure}
\includegraphics[width=0.8\columnwidth]{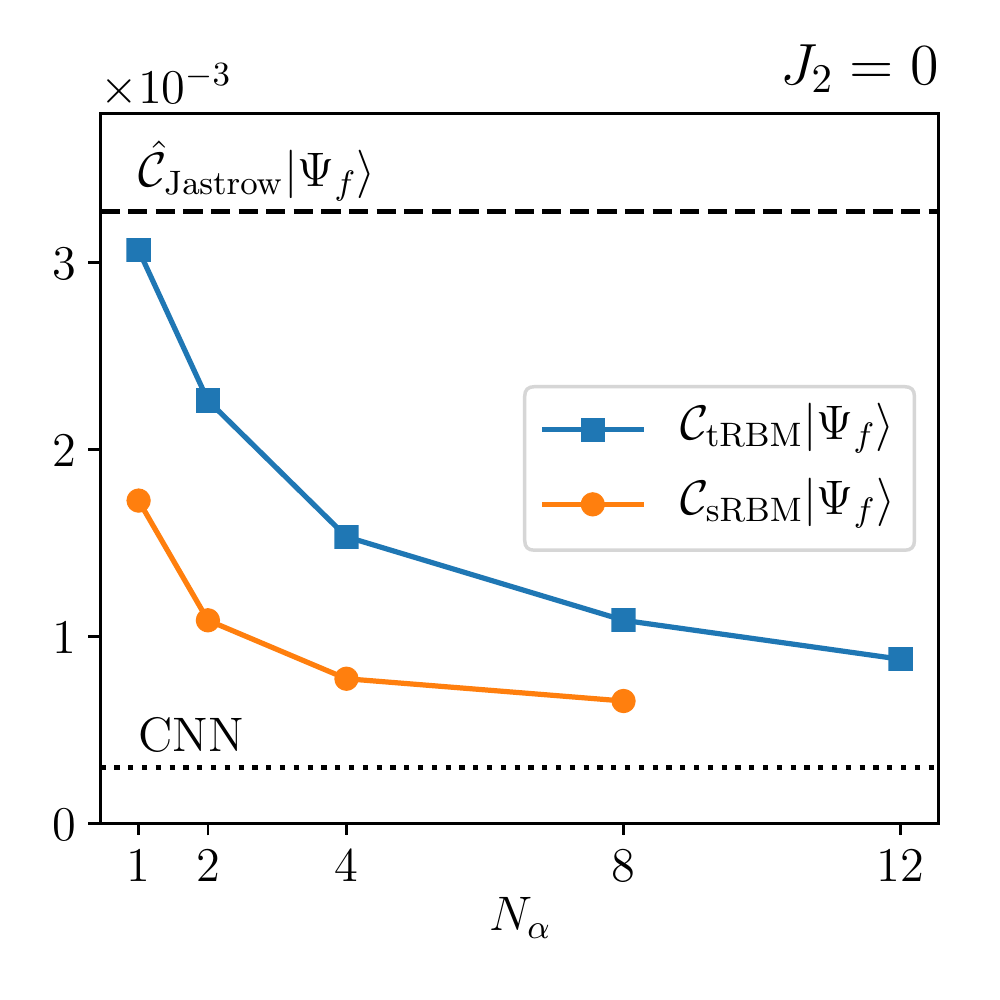}
\caption{\label{fig:energy_heis_10x10}
Relative error of the VMC energies for the Heisenberg model on the $10\times 10$ square lattice, computed with respect to the exact result of quantum
Monte Carlo~\cite{sandvik1997,calandra1998}. The relative error of the RBM-fermionic wave function is plotted as a function of the number of hidden
units: blue squares refer to the case of translationally invariant RBM correlator $\hat{\mathcal{C}}_{\mathrm{tRBM}}$, while orange circles correspond
to the fully symmetric RBM correlator $\hat{\mathcal{C}}_{\mathrm{sRBM}}$. The error bars are smaller than the size of the dots. The dotted line refers
to the relative error of the CNN quantum state of Ref.~\onlinecite{choo2019}.}
\end{figure}

\begin{figure}
\includegraphics[width=\columnwidth]{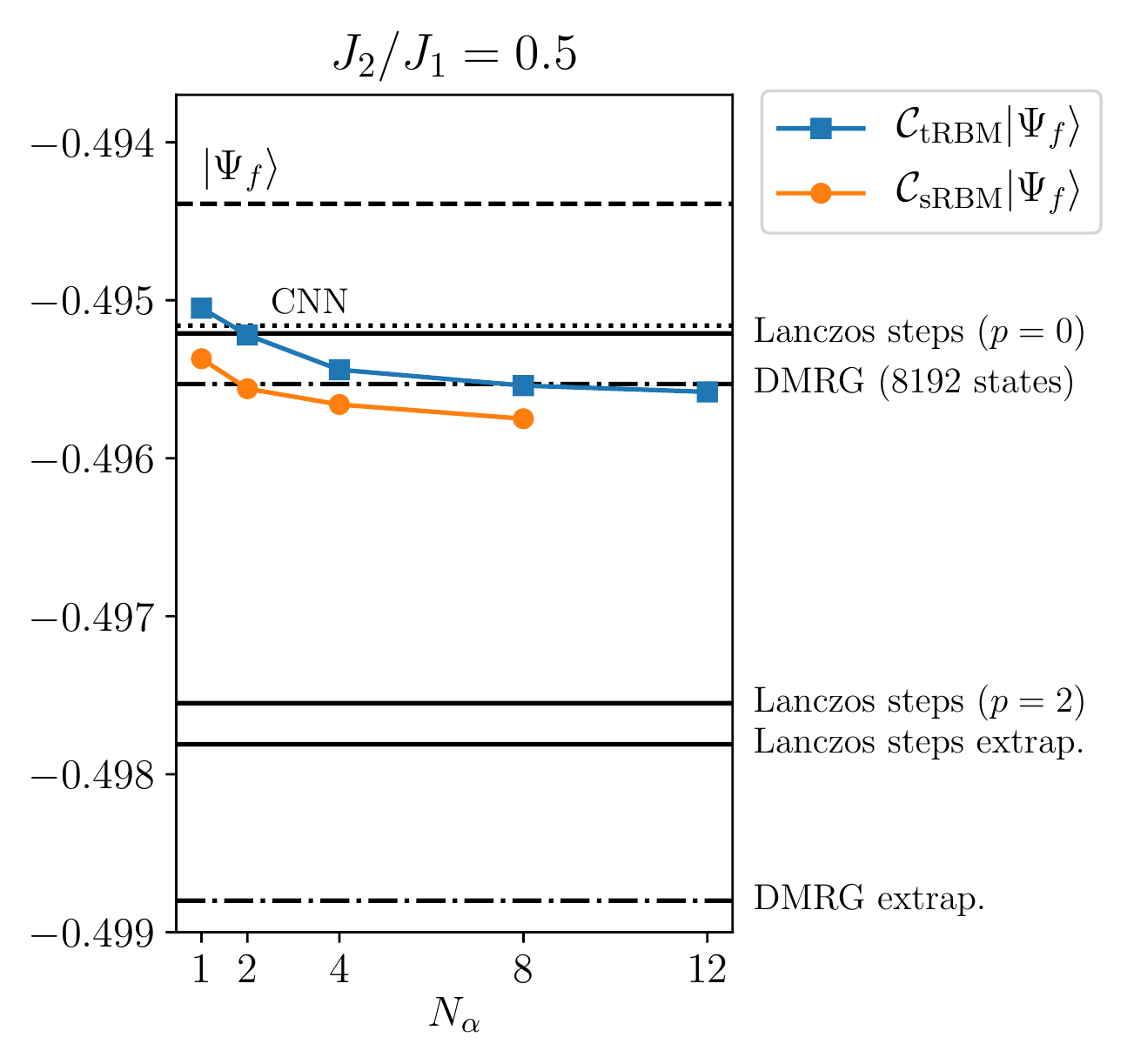}
\caption{\label{fig:energy_j1j2_10x10}
VMC energies for the $J_1-J_2$ model on the $10\times 10$ square lattice in the frustrated regime ($J_2/J_1=0.5$). The variational energies of the
RBM-fermionic wave function are plotted as a function of the number of hidden units: blue squares refer to the case of translationally invariant RBM
correlator $\hat{\mathcal{C}}_{\mathrm{tRBM}}$, while orange circles correspond to the fully symmetric RBM correlator $\hat{\mathcal{C}}_{\mathrm{sRBM}}$.
The error-bars are smaller than the size of the dots. As a comparison we report several different results. The dashed line indicates the energy of the
fermionic wave function of reference. Density-matrix renormalization group (DMRG) energies from Ref.~\onlinecite{gong2014} are plotted with a dotted-dashed
line: the highest energy corresponds to the most accurate result obtained by a DMRG calculation (using $8192$ $SU(2)$ states), while the lowest energy
line corresponds to the value which was obtained by extrapolating DMRG data with respect to the truncation error. Full lines represent the results of
Ref.~\onlinecite{hu2013}, in which Lanczos steps were applied to a fermionic wave function in order to improve its accuracy. Three values are reported
here: the highest energy is obtained with the pure fermionic wave function ($p=0$, i.e., no Lanczos steps), the middle one by the application of two
Lanczos steps ($p=2$), while the lowest one is the result of the variance extrapolation. Finally, the variational energy obtained with the CNN quantum
state of Ref.~\onlinecite{choo2019} is depicted with a dotted line.}
\end{figure}
%%%%%%%%%%%%%%%%%%%%%%%%%%%%%%%%%%%%%%%%%%%%%%%%%%%%%%%%%%%%%%%%%%%%%

\subsection{The $J_1-J_2$ model on the square lattice}

We first discuss the variational Monte Carlo results for the $J_1-J_2$ model on a $6\times 6$ square lattice. For $J_2=0$, the auxiliary fermionic
Hamiltonian $\mathcal{H}_0$ which defines the variational {\it Ansatz} contains a N\'eel magnetic field (with $Q=(\pi,\pi)$) in Eq.~(\ref{eqn:AF}) and
a complex hopping term in Eq.~(\ref{eqn:BCS}), which induces a staggered flux through the square plaquettes~\cite{affleck1988}. This fermionic wave
function possesses the sign structure of the exact ground state of the Heisenberg model, i.e., it follows the so-called Marshall-Peierls sign
rule~\cite{marshall1955,becca2011}. Therefore, it is sufficient to use an RBM correlator with real parameters, so that $C_{\mathrm{RBM}}(\sigma)\ge0$.
On the contrary, in the frustrated regime, the sign structure of the exact ground state is unknown. For $J_2/J_1=0.5$ we combine a complex-valued RBM
correlator and a spin liquid fermionic wave function. The latter is obtained by considering an auxiliary Hamiltonian with a uniform real hopping and a
$d_{x^2-y^2}$ pairing at first neighbors, and a $d_{xy}$ pairing at fifth neighbors~\cite{hu2013}. This wave function corresponds to a $Z_2$ spin liquid
state, which satisfies all the symmetries of the model after the Gutzwiller projection~\cite{wen2002}. Once more, we point out that all the parameters in the
variational state, i.e., the weights and biases of the RBM and the couplings included in $\mathcal{H}_0$, are fully optimized through the stochastic
reconfiguration technique~\cite{sorella2005,becca2017}.

One of the purposes of this work is evaluating the accuracy gain provided by the use of the RBM correlator. For magnetically ordered phases (e.g., $J_2=0$),
the best fermionic state breaks the $SU(2)$ symmetry [since $\Delta_{\rm AF}$ is finite in Eq.~(\ref{eqn:AF})]; in this case, the reference state also
contains the Jastrow factor, since it plays an important role to include quantum fluctuations. Instead, within the non-magnetic phase (e.g., $J_2/J_1=0.5$),
the best fermionic state does not break the $SU(2)$ symmetry (i.e., $\Delta_{\rm AF}=0$) and the Jastrow factor typically gives a negligible contribution.
In this case, we choose to compare the accuracy of the combined RBM-fermionic wave function to the pure Gutzwiller-projected BCS state. In both cases,
the RBM correlator is expected to yield an improvement of the accuracy when all the parameters are properly optimized, since it can be shown that the
Jastrow factor can be represented exactly as a RBM~\cite{nomura2017,glasser2018}. The underlying question we address is understanding to which extent the
application of the RBM correlator improves the accuracy and the physical content of the variational wave function.  For the $J_1-J_2$ model, we consider
both the translationally invariant [$\hat{\mathcal{C}}_{\mathrm{tRBM}}$, Eq.~(\ref{eq:rbm_translations})] and the fully symmetric
[$\hat{\mathcal{C}}_{\mathrm{sRBM}}$, Eq.~(\ref{eq:rbm_allsymm})] RBM correlators.

In Fig.~\ref{fig:energy_j1j2_6x6} we report the relative error of the variational energy of the $J_1-J_2$ model on the $6\times 6$ lattice with respect to
the exact value, obtained by Lanczos diagonalization. This quantity is defined as 
\begin{equation}\label{eq:rel_err_energy}
\Delta E=\left|\frac{E_\mathcal{C}-E_0}{E_0}\right|,
\end{equation}
where $E_\mathcal{C}$ is the energy of a given variational {\it Ansatz} and $E_0$ is the exact ground-state energy. The results clearly show that the variational 
wave function is overall more accurate in the unfrustrated regime with respect to the frustrated one. In particular, at $J_2=0$ the inclusion of the RBM provides 
a large energy gain with respect to the fermionic reference state ($\Delta E \approx 0.4\%$): the relative error of the energy improves by a factor
$\approx 8$ when applying the translationally invariant RBM ($\Delta E\approx0.05 \%$ for $N_\alpha=12$) and a factor of $\approx20$ when applying the fully
symmetric RBM ($\Delta E \approx 0.02 \%$ for $N_\alpha=12$). In general, for each value of $N_\alpha$ we observe that the inclusion of the point group
symmetries in the RBM correlator halves the relative error of the variational energy. By contrast, at $J_2/J_1=0.5$ the accuracy gain is considerably
smaller: the relative error of the energy changes from $\Delta E \approx 0.6\%$ to $\Delta E \approx 0.3\%$ by applying a translationally invariant RBM
correlator, and the addition of point group symmetries is far less effective than what we observe for the unfrustrated case.
In Fig.~\ref{fig:energy_j1j2_6x6} the variational energies obtained with the RBM-fermionic wave functions are compared to the ones of
Ref.~\onlinecite{choo2019}, in which a convolutional neural network (CNN) quantum state is employed: the CNN wave function is more accurate in the
unfrustrated phase, while the RBM-fermionic {\it Ansatz} gives better energies in the frustrated regime.

To further elucidate the ability of the RBM-fermionic wave functions in capturing the ground state properties of the model, we compute the spin-spin
correlations at different distances on the lattice. Whereas the exact ground state wave function of the $J_1-J_2$ model on finite cluster possesses
all the symmetries of its Hamiltonian, most of the variational {\it Ans\"atze} used in our calculations break the spin $SU(2)$ symmetry of the model
due to the presence of the RBM correlator, the Jastrow factor, or the magnetic field $\Delta_{\rm AF}$. Therefore, to investigate the spin symmetry
properties of the resulting states, we separate the computation of the spin-spin correlations at distance $R$ into the in-plane part,
\begin{equation}\label{eq:xycorr}
\mathcal{S}^{xy}_R\equiv
\frac{1}{2N} \sum_{i=1}^N \left( S_i^xS_{i+R}^x + S_i^yS_{i+R}^y\right),
\end{equation}
and the out-of-plane part,
\begin{equation}\label{eq:zcorr}
\mathcal{S}^{z}_R\equiv \frac{1}{N} \sum_{i=1}^N S_i^zS_{i+R}^z.
\end{equation}
We then compare the variational estimates of the two contributions separately with the exact value
${\langle\mathcal{S}_R\rangle_0\equiv1/3 \langle \mathbf{S}_0 \cdot \mathbf{S}_R \rangle_0= \langle S^a_0 S^a_R \rangle_0}$, $a=x,y,z$.
Here $\langle \dots \rangle_0$ indicates the expectation value over the exact ground state $|\Psi_0\rangle$, obtained by Lanczos diagonalization. Thus, 
the relative error of the spin-spin correlations at distance $R$ is computed as 
\begin{equation}\label{eq:rel_err_spinspin}
\Delta {\cal S}^{\alpha}_R = \frac{\langle \mathcal{S}^{\alpha}_R \rangle_{\mathcal{C}}-\langle\mathcal{S}_R\rangle_0}{\langle\mathcal{S}_R\rangle_0},
\end{equation}
where $\langle \dots \rangle_{\mathcal{C}}$ indicates the expectation value over a given variational wave function, and $\alpha=xy,z$ refer to the in-plane, 
Eq.~(\ref{eq:xycorr}), and out-of plane, Eq.~(\ref{eq:zcorr}), estimators.

In Fig.~\ref{fig:corr_heis_6x6_r}, we show the in-plane and out-of-plane variational correlations as a function of distance, in comparison to the exact
value. The results are obtained with the aforementioned RBM-fermionic wave function $\hat{\mathcal{C}}_{\mathrm{sRBM}}|\Psi_f\rangle$ ($N_\alpha=1$) for
$J_2=0$.  Due to the presence of the antiferromagnetic parameter $\Delta_{\rm AF}$, which induces magnetic ordering in the $xy$ plane, the in-plane
correlations overestimate the exact ones in absolute value, while the out-of-plane correlations are underestimated. This tendency is observed for the
spin-spin correlations at any distance and for any value of $N_\alpha$. The results obtained by increasing the number of hidden units are shown in
Fig.~\ref{fig:corr_heis_6x6}, where the relative error of the variational estimates of the correlations with respect to the exact value is reported for
some selected distances $R$. We observe a systematic improvement of the accuracy when the number of hidden units is increased, with both the in-plane
and the out-of-plane correlations approaching the exact value. The fact that these two terms tend to get closer to each other when $N_\alpha$ is
increased indicates that the RBM correlator tries to restore the anticipated spin $SU(2)$ symmetry in the wave function. As expected, since the optimal
wave function is computed by minimizing the ground state energy, the most accurate values for the correlations are obtained at first-neighbors. We note
also that, at any distance, the RBM correlator systematically provides a more accurate estimation of the spin-spin correlations than the simple Jastrow
factor.

Before moving to the frustrated case, we explain the motivation of our choice to consider the accuracy of $\mathcal{S}^{xy}_R$ and $\mathcal{S}^{z}_R$ 
  {\it separately}, instead of using the isotropic spin-spin correlation, $(\mathcal{S}^{z}_R+2\mathcal{S}^{xy}_R)/3$. When performing the optimization of
the variational parameters of the wave function, we observed that the final set of weights and biases of the RBM can depend on their initial values.
Indeed, the parametrization of the RBM displays a considerable degree of redundancy, as indicated by the presence of zero eigenvalues of the covariance
matrix of Eq.~\eqref{eq:Smatrix_def}~\cite{sorella2005,becca2017}. Therefore, the optimization procedure can end into different local minima
which can have equivalent energies and slightly different in-plane and out-of-plane spin-spin correlations. However, this small difference, which is
typically negligible in the unfrustrated regime, can be enhanced or reduced in a random fashion by cancellation of error when the isotropic correlation
is computed. Therefore, we argue that a study of the separate components of the correlation function, rather than the isotropic counterpart, provides a
better characterization of the symmetry properties of our {\it Ansatz}.

At $J_2/J_1=0.5$, the situation is considerably different from the unfrustrated case, as demostrated in Fig.~\ref{fig:corr_j2_6x6}. The accuracy of the
variational correlations does not show a systematic improvement with the number of hidden units. Indeed, even if in general the results are more accurate
for $N_\alpha=12$ than $N_\alpha=1$, the behavior of the relative error is not as smooth as the respective one observed in Fig.~\ref{fig:corr_heis_6x6}
for $J_2=0$. Moreover, in some cases the relative error obtained by applying the RBM correlator is larger than the one obtained from the use of the
simple Jastrow factor. A regular improvement is observed only at first and second neighbors, which are the correlations contributing to the value of the
energy. Most importantly, the role of the RBM correlator regarding the expected $SU(2)$ symmetry is not clear; we find that the out-of-plane correlations 
display a better improvement than the in-plane ones when $N_\alpha$ is increased. We argue that this irregular behaviour of the accuracy of the spin-spin 
correlations is a consequence of the presence of several local minima in the optimization of the variational parameters of the RBM correlator, which lead 
to states with markedly different energies and correlation functions. While the effect is already present in the unfrustrated Heinseberg model, it is 
enhanced in the highly-frustrated regime of the $J_1-J_2$ model. 

To investigate this issue, we performed $40$ distinct optimizations of the $\hat{\mathcal{C}}_{\mathrm{tRBM}}|\Psi_f\rangle$ {\it Ansatz} with $N_\alpha=4$ 
at $J_2/J_1=0.5$, choosing different values of the RBM parameters as starting point. The relative error of the spin-spin correlations obtained by the $40$ 
resulting wave functions is plotted as a function of the relative error of the variational energy in Fig.~\ref{fig:scattered_corr}. We observe that the 
accuracy of the results show considerably large fluctuations. In particular, while the out-of-plane correlations seem to be more accurate when the variational 
energy is lower, an opposite effect is observed for the in-plane terms at some distances. The RBM correlator, which is a function of the $\sigma_z$ degrees 
of freedom, tends to ``sacrifice'' the accuracy of the in-plane correlations for the sake of improving the variational energy. This numerical experiment 
suggests that the complexity of the optimization landscape of RBM in the presence of frustration is significantly different from the unfrustrated case: while 
in the unfrustrated case all the local minima display similar energies and correlation functions, the highly frustrated regime exhibits a wide array of local 
minima with similar variational energy but strinkingly different correlation functions. We speculate that these minima are due to the presence of a possible 
glassy phase in the optimization induced by frustration~\cite{pmlr-v80-baity-jesi18a}.

We conclude our analysis of the $J_1-J_2$ model by presenting the variational energies for the $10\times10$ lattice. The wave functions we employ have
the same form of the ones used for the $6\times 6$ lattice. However, while the number of parameters entering $\mathcal{H}_0$ is the same, the number of
weights of the RBM increases linearly with the size. For the Heinseberg model ($J_2=0$), in Fig.~\ref{fig:energy_heis_10x10} we show the relative error
of the variational energy with respect to the exact one, computed by quantum Monte Carlo~\cite{sandvik1997,calandra1998}. The relative error of the
RBM-fermionic wave function is of the same order of magnitude of the one obtained on the $6\times 6$ lattice and shows a remarkable energy gain with
respect to the Jastrow-fermionic state. However, the accuracy gain provided by the inclusion of the point group symmetries is slightly smaller than
the one observed for the $6\times 6$ lattice. We note that the variational energy of the CNN quantum state of Ref.~\onlinecite{choo2019} is lower than
the best RBM-fermionic energy found in this work. Here, we emphasize the fact that the CNN state employs a larger number of variational parameters
(3838 complex numbers~\cite{choo2019}) than the RBM-fermionic wave function with $N_\alpha=8$ (810 real numbers). However, the local structure of the
CNN is advantageous in the process of optimization, since the optimal parameters obtained for a smaller lattice can be employed as a starting point for
the optimization of the wave function on a larger lattice~\cite{choo2019}. The same procedure cannot be applied in the case of the RBM correlator, due
to its highly nonlocal structure, which implies that the optimization of the parameters of this network necessarily becomes harder when the size of the
system increases.

In Fig.~\ref{fig:energy_j1j2_10x10} we compare our variational energies in the frustrated phase, $J_2/J_1=0.5$, with several different results from
literature. Here, at variance with the unfrustrated case, the variational energies obtained by using the RBM-fermionic wave function are better than the
ones of the CNN of Ref.~\onlinecite{choo2019}, and are very close to the best density matrix renormalization group (DMRG) estimates of
Ref.~\onlinecite{gong2014}. However, a considerably lower variational energy is obtained in Ref.~\onlinecite{hu2013}, where a fermionic wave function,
defined by a BCS Hamiltonian $\mathcal{H}_0$ which contains two additional $d_{x^2-y^2}$ pairings (at fourth- and sixth-neighbors) with respect to the
one employed in this work, is improved by the application of few Lanczos steps. The relative energy gain provided by two Lanczos steps, which require
the addition of only two variational parameter, is remarkably larger than the improvement which is obtained by the application of the RBM correlator,
which contains more than $2000$ parameters for $N_\alpha=12$. In general, as already observed for the $6\times6$ lattice, in the frustrated regime
the RBM correlator yields a much smaller energy gain with respect to the unfrustrated case.

In summary, these results suggest that the RBM correlator provides a systematic way of improving the description of magnetically ordered phases beyond
the Jastrow factor, where the RBM effectivetively induces out-of-plane fluctuations that counterbalance the in-plane magnetic order induced by
$\Delta_{\rm AF}$. In the frustrated regime, even though the application of the RBM leads to better variational energies, it generally does not improve
the description of the correlation functions beyond nearest neighbors. Apart from their numerically expensive training procedure, the RBM's energetic
enhancement comes at the price of breaking of the $SU(2)$ symmetry of the fermionic wave function. The symmetry breaking we observe is especially evident
in the correlation functions beyond nearest neighbors, which do not directly affect the variational energy during the optimization procedure. Fortunately,
all these aspects can be addressed in the future using a recently introduced parametrization of the RBM correlator, which by construction satisfies the
$SU(2)$ symmetry of the models considered in our study~\cite{vieijra2019}.

%%%%%%%%%%%%%%%%%%%%%%%%%%%%%%%%%%%%%%%%%%%%%%%%%%%%%%%%%%%%%%%%%%%%%
\begin{figure}
\includegraphics[width=0.8\columnwidth]{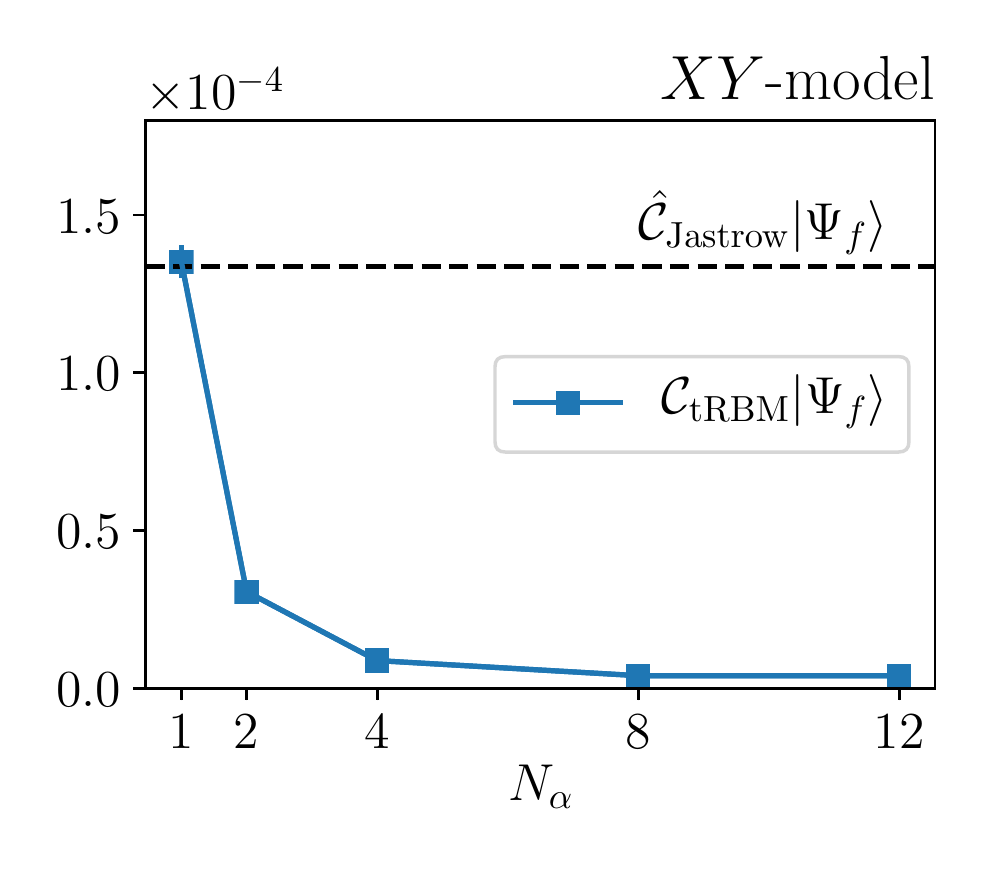}
\caption{\label{fig:xymodel}
Relative error of the VMC energies $\Delta E$ [see Eq.~\eqref{eq:rel_err_energy}] with respect to the exact ones for the $XY$ model on the $6 \times 6$ 
square lattice. The blue squares correspond to the relative error of the RBM-fermionic wave function as a function of the number of hidden units ($N_\alpha$). 
The error-bars are smaller than the size of the dots. The dashed line represents the relative error of the fermionic wave function of reference, which includes 
a Jastrow factor.}
\end{figure}
%%%%%%%%%%%%%%%%%%%%%%%%%%%%%%%%%%%%%%%%%%%%%%%%%%%%%%%%%%%%%%%%%%%%%

\subsection{The $XY$ model on the square lattice}

As already pointed out, one of the main drawbacks of the introduction of the RBM correlator is the breaking of the spin $SU(2)$ symmetry of the wave
function. For this reason, we evaluate the accuracy of the RBM-fermionic construction for a model whose Hamiltonian has lower symmetry, i.e., $XY$ model
of Eq.~(\ref{eq:xymodel}). The exact ground state of the $XY$ model has the same sign structure of the one of the Heinseberg model, i.e. it follows
the Marshall-Peierls rule~\cite{marshall1955}. Thus, we employ an analogous RBM-fermionic wave function like the one used for the Heinseberg model 
(with real weights and biases). The variational results are reported in Fig.~\ref{fig:xymodel} for the translationally invariant correlator
$\hat{\mathcal{C}}_{\mathrm{tRBM}}$. The relative error of the variational energy with respect to the exact one is at least a factor of $10$ smaller
than the one obtained for the Heisenberg model and the accuracy gain provided by the RBM correlator is remarkable ($\Delta E\approx0.0004\%$ for 
$N_\alpha=12$). The higher accuracy of the wave function is related to the fact that here the  symmetry of the variational {\it Ansatz} is consistent 
with the spin symmetry of the model.

%%%%%%%%%%%%%%%%%%%%%%%%%%%%%%%%%%%%%%%%%%%%%%%%%%%%%%%%%%%%%%%%%%%%%
\begin{figure}
\includegraphics[width=\columnwidth]{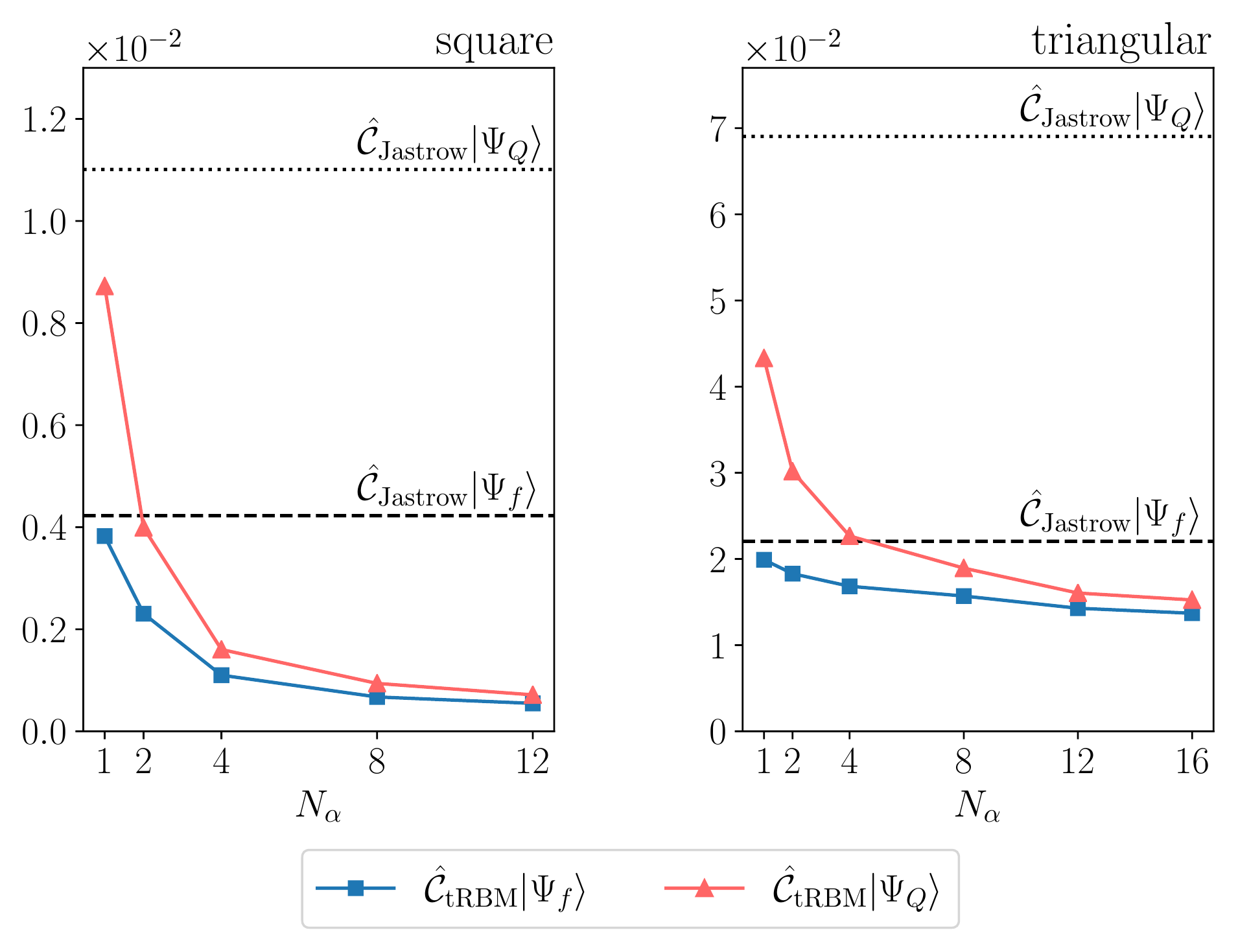}
\caption{\label{fig:triang_vs_square}
Relative error of the VMC energies $\Delta E$ [see Eq.~\eqref{eq:rel_err_energy}] with respect to the exact ones for the Heisenberg model on the $6\times 6$ 
square (left) and triangular (right) lattices. The blue squares (pink triangles) correspond to the relative error of the wave function obtained by applying 
a translationally invariant RBM correlator to $|\Psi_f\rangle$ ($|\Psi_Q\rangle$), as a function of the number of hidden units $N_\alpha$. The error-bars
are smaller than the size of the dots. The dashed lines represent the relative error of the Jastrow-fermionic wave function
$\hat{\mathcal{C}}_\mathrm{Jastrow}|\Psi_f\rangle$, while the dotted ones correspond to  $\hat{\mathcal{C}}_\mathrm{Jastrow}|\Psi_Q\rangle$.}
\end{figure}

\begin{figure}
\includegraphics[width=\columnwidth]{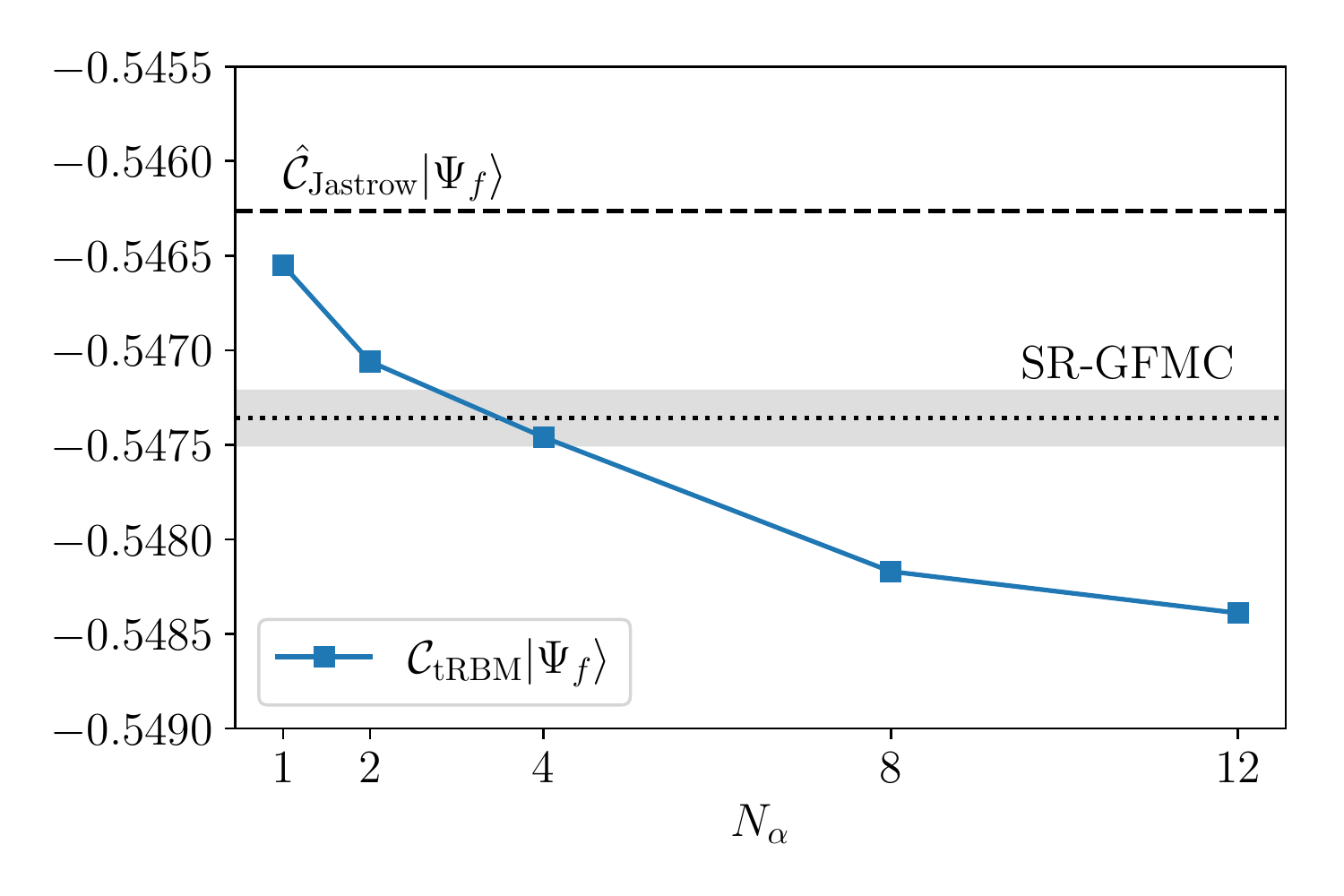}
\caption{\label{fig:12x12_triang}
VMC energies for the Heisenberg model on the $12\times 12$ triangular lattice. The variational energies of the RBM-fermionic wave function are plotted
as a function of the number of hidden units (blue squares). The error bars are smaller than the size of the dots. The dashed line indicates the energy
of the Jastrow-fermionic wave function of reference. The dotted line corresponds to the Green's function Monte Carlo result of
Ref.~\onlinecite{capriotti1999}, whose error bar is represented by the grey shaded area.}
\end{figure}
%%%%%%%%%%%%%%%%%%%%%%%%%%%%%%%%%%%%%%%%%%%%%%%%%%%%%%%%%%%%%%%%%%%%%

\subsection{The Heisenberg model on the triangular lattice}

Our previous results suggest the idea that the application of the RBM correlator is more effective for magnetically ordered phases, as exemplified in the
Heisenberg and $XY$ model, rather than for non-magnetic ones, as in the frustrated region of the $J_1-J_2$ model. However, in the N\'eel phases considered
above, the exact sign structure of the wave function is particularly simple and exactly captured by the fermionic part of the variational {\it Ansatz}.
To try to disentangle whether the successes observed in our simulations are related to the special structure of the sign or to the presence of magnetic order, we
consider a model whose ground state is magnetically ordered but displays a non-trivial sign structure: the Heisenberg model on the triangular lattice.
In this case, the fermionic part of the {\it Ansatz} is constructed via  an auxiliary Hamiltonian $\mathcal{H}_0$ which features a magnetic field
$\Delta_{\mathrm{AF}}$ with pitch vector $Q=(\frac{4\pi}{3},0)$ [or, equivalently, $Q=(\frac{2\pi}{3},\frac{2\pi}{\sqrt{3}})$] and a real nearest-neighbor
hopping $t$. The sign structure of the hopping generates an alternation of $0$ and $\pi$ fluxes threading the triangular plaquettes~\cite{iqbal2016}.
While in square lattice a real parametrization on top of the fermionic state already gives an accurate representation for both signs and amplitudes of
the exact ground state, we anticipate that the triangular lattice Heisenberg model requires a complex-valued correlator to approximate the unknown sign
structure of the wave function induced by the geometric frustration of the problem. Thus, on top of the fermionic state, we apply a translationally
invariant RBM correlator with complex parameters ($\hat{\mathcal{C}}_{\mathrm{tRBM}}$).

In Fig.~\ref{fig:triang_vs_square}, we compare the accuracy of the RBM-fermionic wave function for the Heisenberg model on the $6\times6$ square and
triangular lattices. The energy gain provided by the application of the RBM correlator is considerably larger in the case of the square lattice, where
the relative error of the energy decreases of a factor $\approx 8$, with respect to the case of the triangular lattice, where it decreases of a factor
$\approx 1.5$ (from $\Delta E \approx 2.2\%$ to $\Delta E \approx 1.4\%$). Overall, the variational energy is more accurate on the square lattice than
on the triangular lattice, and the relative errors differ by an order of magnitude.

In Fig.~\ref{fig:triang_vs_square}, we also compare the results of the aforementioned RBM-fermionic wave functions to the ones obtained by simpler
{\it Ans\"atze}, which are constructed by setting the hopping terms to zero and considering an auxiliary fermionic Hamiltonian with only magnetic field
($\mathcal{H}_0=\mathcal{H}_{\mathrm{AF}}$). In this way, the fermionic degrees of freedom are localized and $|\Psi_{f}\rangle$ reduces to a (projected)
product state $|\Psi_{Q}\rangle=\mathcal{P}_{S^z_{tot}=0}\prod_{i=1}^N (|\uparrow\rangle_i+e^{iQR_i}|\downarrow\rangle_i)$, which displays ``classical''
order in the $xy$-plane. This wave function can be employed as a reference state for the application of a Jastrow factor~\cite{carrasquilla2013} or the
RBM correlator. Here, we apply a translationally invariant RBM correlator to $|\Psi_{Q}\rangle$. We observe that the energy gain provided by the
presence of the hopping term in the fermionic {\it Ansatz} is remarkably large when the simple Jastrow factor is applied to the reference state. However,
when the RBM correlator is employed, the contribution of the hopping term becomes less important, decreasing considerably with the number of hidden units,
which suggests that the RBM replaces the effect of the fermionic hopping term in the state.

Finally, in Fig.~\ref{fig:12x12_triang} we present the variational energies obtained for the Heisenberg model on the $12\times12$ triangular lattice.
For a large enough $N_\alpha$, the variational energy of the RBM-fermionic wave function is more accurate than the Green's function Monte Carlo (GFMC)
result of Ref.~\onlinecite{capriotti1999}. To summarize our numerical experiments, we surmise that the high accuracy of the results for the unfrustrated 
square lattice Heisenberg model is due to having an exact representation of the sign structure built in the Ansatz, which in turn, alleviates the energy 
optimization problem. On the other hand, the lower accuracy in the triangular lattice model is presumably due to a combination of the approximate nature 
of the sign structure imposed by our Ansatz and the energy optimization problem in the presence of complex numbers. Since magnetically ordered states are 
pervasive in frustrated magnetism, it remains an important issue to unequivocally establish whether the origin of the high accuracy of the results for 
the Heisenberg model on the square lattice is only due to the absence of frustration or to the fact that the sign structure is exactly known in the 
unfrustrated case.

\section{Conclusion}\label{sec6}

Inspired by Hinton's product of experts idea~\cite{hinton2002}, we have studied a neural augmentation of the parton construction that combines a family
of physically motivated Gutzwiller-projected variational states with a complex-valued RBM correlator. We focused our attention on various prototypical
spin models traditionally used in condensed matter physics which exhibit a wide array of conventional ground states with magnetic order, as well as
more exotic ones where a spin-liquid behavior has been anticipated.

In agreement with previous results based upon neural networks alone~\cite{carleo2017,choo2019}, our calculations showed that RBMs are very effective in the
unfrustrated Heisenberg and $XY$ models on the square lattice, where the knowledge of the exact sign structure of the ground state allows us to use a real
parametrization of the RBM. Here, a few hidden units in the neural network are sufficient to reach a striking accuracy, which for the Heisenberg case is 
comparable with the best variational wave functions defined within the bosonic resonating valence-bond picture~\cite{liang1988}. Moreover, we emphasize the 
remarkable ability of the RBM to systematically recover the spin $SU(2)$ symmetry upon increasing the number of hidden units. This is clear not only by 
looking at first-neighbor correlations, but also at further distances: both out-of-plane correlations (i.e., along $z$) and in-plane ones (i.e., in the $xy$ 
plane) exhibit a clear tendendcy to converge toward the exact (isotropic) result upon increasing $N_{\alpha}$.

In the highly frustrated regime, the exact ground-state sign is not known {\it a priori}, even when the system displays magnetic order, like in the Heisenberg 
model on the triangular lattice. Here, we find a substantial energy gain with respect to the original parton wave function, although the variational procedure 
does not yield the same accuray as in the square lattice, even for a relatively large number of hidden units. A similar effect is also observed in the 
highly-frustated regime of the $J_1-J_2$ Heisenberg model, where for $J_2/J_1=0.5$ our neural Gutzwiller-projected wave functions exhibit accuracies beyond 
recent neural network calculations based on CNNs~\cite{choo2019}, though with a lower accuracy than in the $J_2=0$ case. Based on the numerical experiments 
presented in Fig.~\ref{fig:scattered_corr}, we surmise that the application of neural network variational states to frustrated quantum spin systems requires 
an extensive investigation of the intricate relation between the representation power of neural networks to capture highly-entangled states of matter with a 
complicated sign structure and the complexity of the optimization landscape of the problem induced by frustration. Motivated by Ref.~\onlinecite{day2019}, 
the disentangling of these factors could be approached through a clustering analysis of the trained RBM parameters and their associated spin-spin correlation 
functions, which may shed light onto the interplay between frustration, sign structure and entanglement~\cite{grover2015}, and the rough optimization landscape 
of the problem~\cite{pmlr-v80-baity-jesi18a}.

At variance with the unfrustrated limit, the $SU(2)$ symmetry is broken and hardly recovered when increasing $N_{\alpha}$ in the frustrated case: the
out-of-plane correlations are clearly more accurate than in-plane ones, highligthing the ``asymmetry'' of the RBM form, which is defined in terms of
the $z$-component of the local spin. In this regard, our calculations  provide a first investigation into the {\it inductive bias} of the RBM applied
to ground states of frustrated systems, where we observe a strong tendency to break $SU(2)$ symmetry in our setting. Thus, the recent proposal to generalize
the RBM to fulfill the $SU(2)$ symmetry~\cite{vieijra2019} may pave the way for studying fully symmetric neural Gutzwiller-projected wave functions, which will 
be especially relevant in the understanding of models for spin liquid phases and other states exhibiting fractionalized excitations and gauge structures.

\acknowledgements

We would like to thank K. Choo, G. Carleo, G. Torlai, and R. Melko for useful discussions. F.F. acknowledges the kind hospitality and financial support
of the Vector Institute for Artificial Intelligence in Toronto, where this project started. We also thank the Kavli Institute for Theoretical Physics
(KITP) in Santa Barbara and the program ``Machine Learning for Quantum Many-Body Physics''.  This research was supported in part by the National Science
Foundation under Grant No. NSF PHY-1748958. J.C. acknowledges support from the Natural Sciences and Engineering Research Council of Canada (NSERC) and 
the Canada CIFAR AI chair program.

%\bibliography{Biblio}


\end{document}